\documentclass[conference]{IEEEtran}
\IEEEoverridecommandlockouts
\usepackage{cite}
\usepackage{amsmath,amssymb,amsfonts}
\usepackage{graphicx}
\usepackage{textcomp}
\usepackage{xcolor}
\usepackage{algorithm}
\usepackage[noend]{algpseudocode}

\makeatletter
\newcommand{\linebreakand}{%
  \end{@IEEEauthorhalign}
  \hfill\mbox{}\par
  \mbox{}\hfill\begin{@IEEEauthorhalign}
}
\makeatother

\def\BibTeX{{\rm B\kern-.05em{\sc i\kern-.025em b}\kern-.08em
    T\kern-.1667em\lower.7ex\hbox{E}\kern-.125emX}}
\begin{document}

\title{Entanglement Routing over Networks with Time Multiplexed Repeaters
\thanks{This work was funded by NSF grant CNS-1955834, NSF-ERC Center for Quantum Networks grant EEC-1941583, and NWO QSC grant BGR2 17.269.}
}

\author{\IEEEauthorblockN{Emily A. Van Milligen}
\IEEEauthorblockA{
\textit{University of Arizona}\\
Tucson, USA \\
evanmilligen@arizona.edu}
\and
\IEEEauthorblockN{Eliana Jacobson}
\IEEEauthorblockA{\textit{University of Arizona}\\
Tucson, USA}
\and
\IEEEauthorblockN{Ashlesha Patil}
\IEEEauthorblockA{
\textit{University of Arizona}\\
Tucson, USA}

\linebreakand
\IEEEauthorblockN{Gayane Vardoyan}
\IEEEauthorblockA{
\textit{University of Massachusetts Amherst}\\
Amherst, USA \\
gvardoyan@umass.edu}
\and
\IEEEauthorblockN{Don Towsley}
\IEEEauthorblockA{\textit{University of Massachusetts Amherst}\\
Amherst, USA \\
towsley@cs.umass.edu}
\and
\IEEEauthorblockN{Saikat Guha}
\IEEEauthorblockA{\textit{University of Maryland}\\
College Park, USA \\
saikat@umd.edu}
}

\maketitle

\begin{abstract}
Quantum networks will enable long-distance entanglement distribution through repeater nodes capable of both generating external Bell pairs with their neighbors—independently and identically distributed (i.i.d.) with probability p—and performing internal Bell State Measurements (BSMs), which succeed with probability q. These probabilities depend on the specific experimental parameters of the network. While global link state knowledge can maximize entanglement generation rates between any two consumers, it is often impractical due to the network’s dynamic nature.

This work evaluates a multipath routing protocol based on local link state knowledge, designed for time-multiplexed repeaters that can perform BSMs across different time steps. We employ simulations across various network topologies to characterize how consumer placement and network topology influence the utility of dynamic and static routing strategies. Results show that increasing the time multiplexing block length, k, enhances the average entanglement generation rate, though at the cost of higher initial latency. When a step-function memory decoherence model is introduced—where qubits are stored in quantum memory for a time that follows an exponential distribution—an optimal k emerges. As p decreases or the average memory lifetime increases, this optimal k grows, reflecting a trade-off where the advantages of time multiplexing must be balanced against the heightened risk of losing previously established entangled pairs.
\end{abstract}


\section{Introduction}
\label{sec:introduction}
Quantum networks will have the ability to generate, distribute, and process quantum information along with classical data. These networks will connect quantum computers, sensors, simulators, and other quantum processors, and enable consumers to share entangled connections on demand over potentially large distances. There are various applications that leverage entangled qubits shared over quantum networks, such as provably-secure communication \cite{Ekert1992,bennett_quantum_1992}, entanglement-enhanced sensing \cite{xia_demonstration_2020,grace_identifying_2021,humphreys_quantum_2013,proctor_multiparameter_2018,zhuang_distributed_2018}, and distributed quantum computing \cite{van_meter_path_2016}. To enable these, it is important to be able to reliably and quickly transmit quantum bits from one point in a network to another in order to establish entanglement between distant users.

A quantum network is comprised of two main components: \textit{nodes}, which are equipped with entanglement sources, quantum memories, quantum processors, and the ability to classically communicate, and \textit{edges}, which connect neighboring nodes via channels that transmit entangled (typically photonic) qubits, such as optical fibers. The nodes can be consumers wanting to utilize entanglement for some application or quantum repeaters, which are fundamental for the successful distribution of entanglement across large distances. The rate that entangled photons can be delivered across an optical fiber decays exponentially with the distance between communicating parties \cite{pirandola_fundamental_2017}. However, when utilizing intermediate repeater nodes, the magnitude of the exponent can be decreased by breaking up the total distance between consumers, which helps to mitigate the loss incurred from the transmitting medium \cite{takeoka_fundamental_2014,guha_rate-loss_2015,pant_rate-distance_2017,duan_long-distance_2001}.

Quantum \textit{links} are maximally-entangled two-qubit states called Bell states shared along edges when two neighboring nodes each successfully store one half of a Bell pair in their quantum memories. The generation of these links has a probability $p$ of succeeding. This value is determined by the actual physical hardware and encompasses optical loss of the link, as well as detector and memory inefficiencies. Repeaters in this paper are quantum switches \cite{lee_quantum_2022,pant_rate-distance_2017}, which can simultaneously attempt to form entangled links with neighboring nodes and dynamically choose on which links to perform entanglement \textit{swaps} by using local network information. A swap is most commonly implemented by performing a Bell state measurement (BSM) on qubits held within a repeater with a probability of success denoted by $q$. If successful, this results in the corresponding qubits held at the repeater’s neighbors to become entangled. However, if the measurement fails, both of the Bell states the repeater shared with its neighbors are lost.

\begin{figure}
    \centering
\includegraphics[width=.45\textwidth]{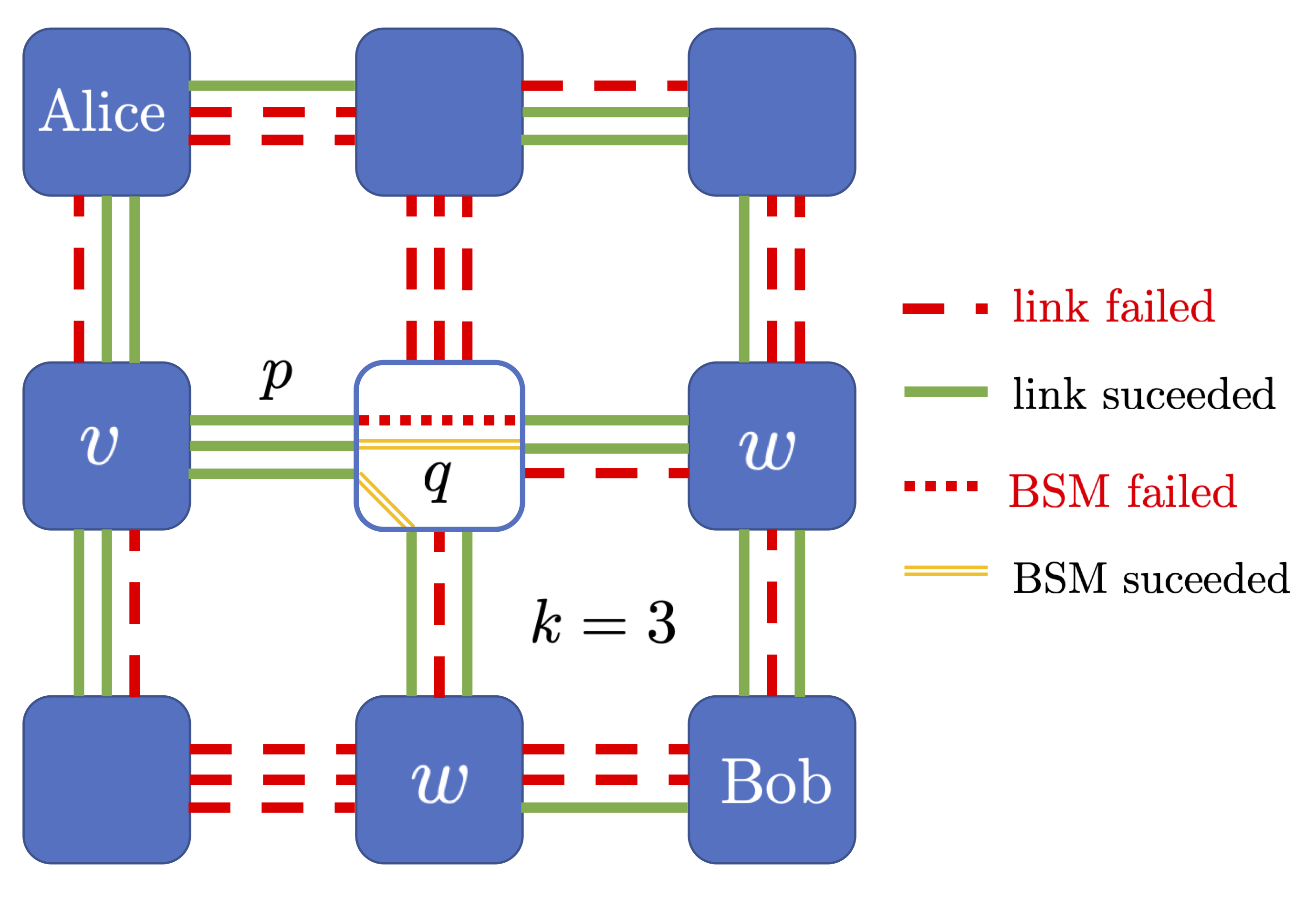}
    \caption{A snapshot of the distance-based routing protocol on a 3 by 3 grid network for $k=3$. The outcomes of the external phase are all depicted, while the center repeater node highlights the decisions made during the internal phase. The repeater node labeled $v$ is the neighbor that shares a Bell pair with the center node, that is closest to Alice, while both of the nodes labeled $w$ are closest to Bob. Although only the decisions of the center node are shown here, all repeater nodes are simultaneously making swapping decisions. }
    \label{fig:ProtocolExplained}
\end{figure}

Increasing the rate of entanglement distribution for networks under different constraints is crucial to the usefulness of quantum networks. The probabilistic nature of link generation means that different links will be present or absent when the network is observed at different instances in time. We refer to each such network state observation as a ``snapshot". Global link state knowledge is needed by a protocol to ensure that two consumers, Alice and Bob, share the maximum number of Bell pairs possible given the initial snapshot. However, global link state knowledge is often not possible as quantum networks can spread out across large distances giving rise to large latencies due to classical communication. Quantum memories cannot support this condition as they have very short storage lifetimes. 

Routing protocols have been proposed to lower latency and therefore the requirements on quantum memories by creating virtual graphs of entangled links before or at the time consumers request entangled pairs \cite{schoute_shortcuts_2016,chakraborty_distributed_2019}. To further reduce this latency, \cite{pant_routing_2019} created a protocol that combined \textit{local link state knowledge} and \textit{multipath routing} to distribute Bell pairs to consumers. It was seen to provide rates larger than those achievable by a single linear repeater chain and only required that each repeater knew the physical topology of the network and the outcomes of their own entanglement generation attempts. Further exploration of distance-based routing schemes can be found in \cite{nguyen_multiple-entanglement_2022,zhang_multipath_2022,li_effective_2021}. Distillation protocols and routing were analyzed in \cite{victora_purification_2020} to maximize the distillable entanglement between consumers. Multipartite entanglement distribution schemes have been investigated in \cite{patil_entanglement_2022, sutcliffe2023multiuser, gyongyosi_decentralized_2018}.

Time multiplexed link generation increases the rate of distribution of entangled qubits. It has been shown that time-multiplexed repeater nodes, equipped with the ability to perform BSMs across different time steps, allow for a sub-exponential decay with distance \cite{dhara_subexponential_2021}. This result is otherwise unattainable with just spatial or spectral multiplexing. Time multiplexing has also been combined with GHZ-projective measurements \cite{Patil_2021}. 

In this paper, we combine the benefits of multipath routing and local link state knowledge with time-multiplexed repeaters to further improve the rate of entanglement distribution.  These time multiplexed repeater nodes are able to hold onto multiple links before routing decisions are made. This increases the likelihood that two repeaters are connected and the number of end-to-end links consumers can generate. We develop two pragmatic routing protocols assuming the repeaters know the network topology and have local-link state knowledge. The \textit{static protocol} uses fixed path routing. It predetermines the pair of neighbors (if any) a repeater connects by performing BSMs. The \textit{dynamic protocol}, on the other hand, uses distance-based metrics to decide which pairs of neighbors are connected via BSMs at a repeater. 

Both protocols are presented in Section II.  In Section III, we will evaluate these protocols on networks where memories have infinite coherence time. In this case, time multiplexing only serves to improve the average rate of the distribution of Bell pairs. We discuss the impact that quantum network structure and consumer locations have on the performance on both of our protocols. In Section IV we reduce the coherence time of our quantum memories, and show that there exists an optimal amount of time multiplexing dependent on the conditions of the network. Last, we offer concluding thoughts in Section V and comment on future directions.

\section{\label{sec:ProtocolDesign}Protocol Design: Multipath Routing with Time Multiplexed Repeaters}
We propose an algorithm that combines local link state knowledge and multipath routing with time multiplexed repeaters. The proposed protocol can be divided into two phases. 

\paragraph*{The External Phase} is the first phase, which lasts for $k$ timesteps, where $k$ is the time multiplexing block length. These timesteps are of length $\tau$ seconds. In each timestep, an attempt is made to generate a pair of entangled qubits between neighboring repeaters along all edges. (In the case where the network has a grid topology, each node attempts this $k$ times with each of its four neighbors.) The probability that the initial link generation succeeds is denoted by $p$. We will refer to the network state after this phase as a snapshot. 

The heralding of the initial success/failure of these Bell pairs is received by neighbors $L/c$ seconds after the entanglement is attempted, where $L$ is the distance between neighbors and $c$ is the speed of light. This initial latency between the creation of the Bell pair and when the pair can actually be utilized exists even when there is no time multiplexing. For this paper, we will assume that each  repeater node has quantum memory buffer greater than $\lceil L/c\tau\rceil+k$ per edge to accommodate this latency as well as that which arises from the time multiplexing. Although this work does not specify a specific type of architecture for the network, the quality of entanglement may degrade during this time. This paper employs the step-function decoherence model used in \cite{Patil_2021,nain_analysis_2020,vardoyan_stochastic_2019} where qubits are held perfectly in memory until some random time, drawn from an exponential distribution with mean lifetime $\mu$, after which the entanglement is discarded. It is assumed that repeaters know when entanglement is discarded, so that the remaining entanglement in the network is not lost. This model is a conservative approximation for memory architectures that have exponentially decaying qubit fidelities with measurable time constants, such as trapped ions and color centers\cite{wang_single_2021}. Initially, we will set $\mu=\infty$. This condition will later be relaxed, so that not all the original Bell pairs are recoverable after waiting the additional $k\tau$ seconds.

\paragraph*{The Internal Phase} is the second phase in which entanglement swaps are attempted within each repeater node with success probability $q$. It is assumed that this process is instantaneous when compared to the first phase. Repeaters can choose different pairs of successful links to connect depending on the protocol being used. This paper investigates two different categories of protocols: \textit{static} and \textit{dynamic}.

Static refers to a fixed path routing protocol that is predetermined based on the physical structure of the underlying network. Let $\theta$ be the number of edge disjoint paths connecting consumers. In the case of regular lattices, this is equivalent to the node degree. A greedy algorithm is used to find the first $\theta$ edge disjoint shortest paths between consumers, Alice and Bob. This information is communicated to the repeaters along these paths, and only connections along these paths will be made. In order to most efficiently use any successfully generated entangled links along edges, a predetermined ordering can be used to maximize the number of Bell pairs at the end. An example of this would be an ordering based on the time slot the links were successfully created, and performing BSMs on the most recent ones on each edge.

Dynamic, or distance based routing, refers to when all repeater nodes simultaneously choose which internal swaps to make using their knowledge of the successful outcomes of the first phase, along with knowledge of their neighbors' relative distances from the consumers, Alice and Bob. A detailed explanation of this process is described below, and pseudo-code formulation is included in Appendix \ref{psuedocode}. We have included a cartoon depiction of this protocol being implemented in Fig.~\ref{fig:ProtocolExplained} for further clarity.

\paragraph*{Distance Based Routing}
Let $n$ be the current node and let $s$ be the number of successful external links at $n$ that have not been swapped yet. As long as there are $s>1$ successful links remaining, swaps continue to occur. Out of the neighbors that are successfully connected to node $n$, let $v$ be the neighbor closest to Alice and define their relative distance to be $d_{Av}$. Let $w$ be the neighbor closest to Bob with a relative distance be $d_{Bw}$. There are many different ways of setting this distance metric, such as Euclidean distance, hop distance etc., which will be specified for the given example.

If $v$ and $w$ are different neighbors, an entanglement swap is attempted. If $v$ and $w$ refer to the same neighbor, then the second closest neighbor to Alice, $v'$, and the second closest neighbor to Bob, $w'$, are found. The quantities $d_{Av'}+d_{Bw}$ and $d_{Av}+d_{Bw'}$ are compared, and the pair that minimizes this choice is chosen to perform a swap. If these quantities are the same, the quantities $d_{Bv'}+d_{Aw}$ and $d_{Bv}+d_{Aw'}$ are compared, and the pair that maximizes this value is chosen to perform a swap. If neither $v'$ nor $w'$ exist, a self connection will be made when possible such that two links between the node and the same neighbor from different timesteps will attempt swapping. (This prevents the protocol from creating holes in the path between consumers and is done as a last resort.) In all of these cases, the procedure is repeated until $s\le1$. 

\paragraph*{Performance Metric} This paper aims to compare the average entanglement distribution rate achieved by different routing protocols over various network conditions. After each entanglement generation attempt occurs, the network will have different links present/absent. This instance of the network can be thought of as a particular network snapshot. The average number of Bell pairs shared between consumers given by a certain protocol for a particular network snapshot is defined as:
\begin{align}
    N(S):=\sum_{l\in\mathcal{L}}\prod_{\substack{i \in l}}q_i
\end{align}
where $\mathcal{L}$ is the set of link-disjoint paths connecting Alice and Bob allowed by the particular snapshot $S$ and by the protocol being evaluated, $l\in\mathcal{L}$ is a single path in $\mathcal{L}$, $i\in l$ is a node along the path $l$ (not including Alice and Bob), and $q_i$ is the probability that a BSM will succeed at node $i$. The average rate of a protocol as a function of $k$ given a certain network is defined as:
\begin{align}
    R(k):=\frac{1}{k}\sum_{S\in\mathcal{S}}P(S,k)N(S)
\end{align}
where $\mathcal{S}$ gives the set of all possible snapshots for a given network, and $P(S,k)$ gives the probability of the particular snapshot $S$ after $k$ timesteps. In this paper, the average rate will be approximated by sampling from Monte Carlo simulations. To compare across different circumstances, we will define our notation such that the average rate is given by $R_{k}(p,q,\mu)$ Bell pairs per time slot, where $k,\ p,\ q,$ and $\mu$ have all been previously defined.

\section{\label{sec:Results with perfect quantum memories}Memories with infinite storage lifetime}
In this section, we evaluate the performance of the dynamic routing protocol under the assumption of perfect quantum memories (\(\mu = \infty\)). We begin with a small six-node network, using hop distance as the metric to inform local swapping decisions. The resulting entanglement distribution rates are compared against those obtained using global link-state knowledge, providing a baseline for quantifying the performance gap attributable to limited local information. We then extend the analysis to a large two-dimensional square grid, where Euclidean distance has been found to yield the highest average entanglement rates. These simulations are used to characterize the behavior of the protocol (dynamic and static versions) in larger, more connected network topologies.  

\subsection{\label{Small-Networks}Small Network Examples}

\begin{figure}
    \centering
\includegraphics[width=.45\textwidth]{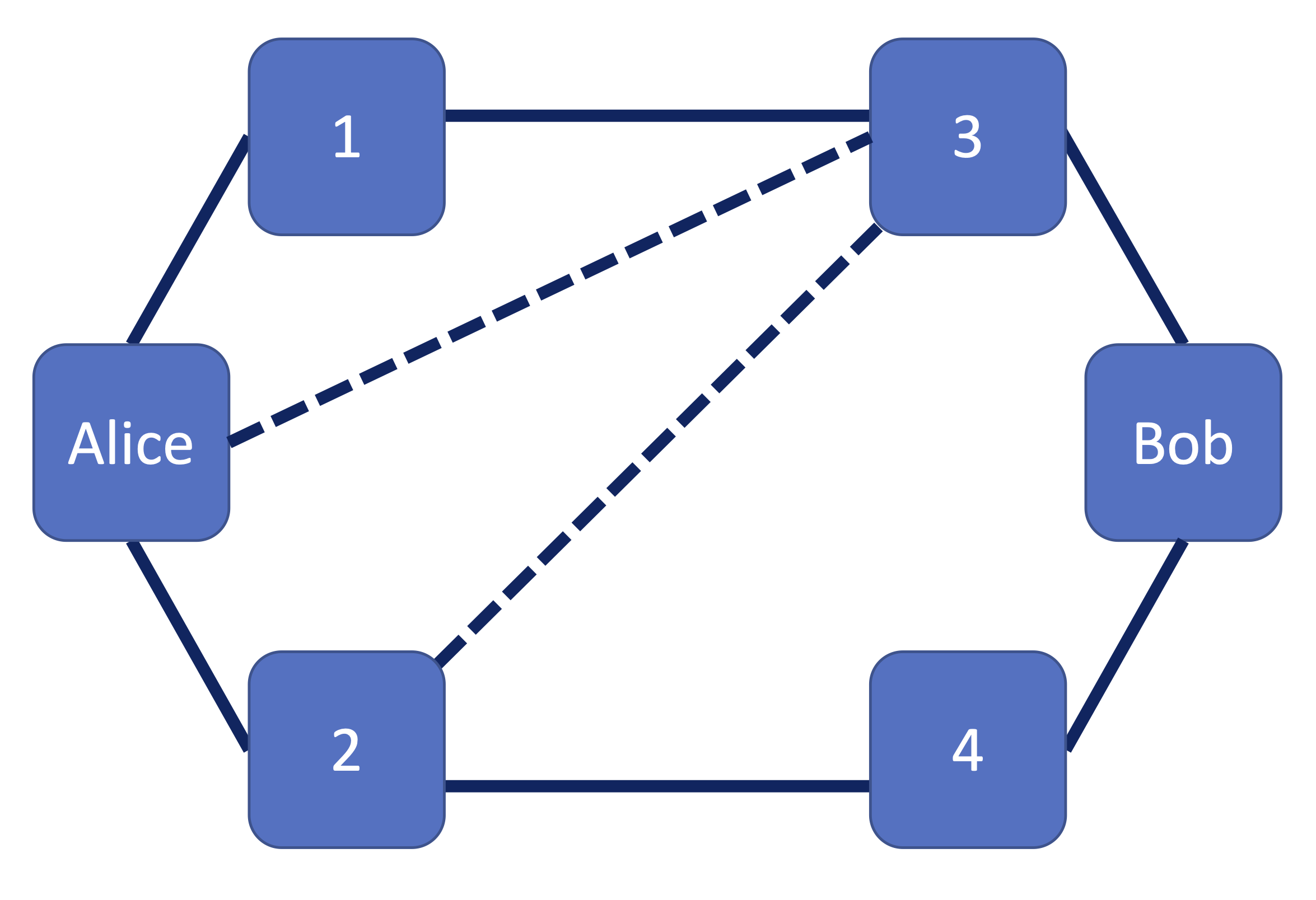}
    \caption{The architecture of our six node graph. The lines depict channels over which links can be generated. The dashed lines are to distinguish the channels between (Alice, 3) and (2, 3) which may be removed to compare the impact of topology.}
    \label{fig:6node}
\end{figure}

\begin{figure}
    \centering
\includegraphics[width=.45\textwidth]{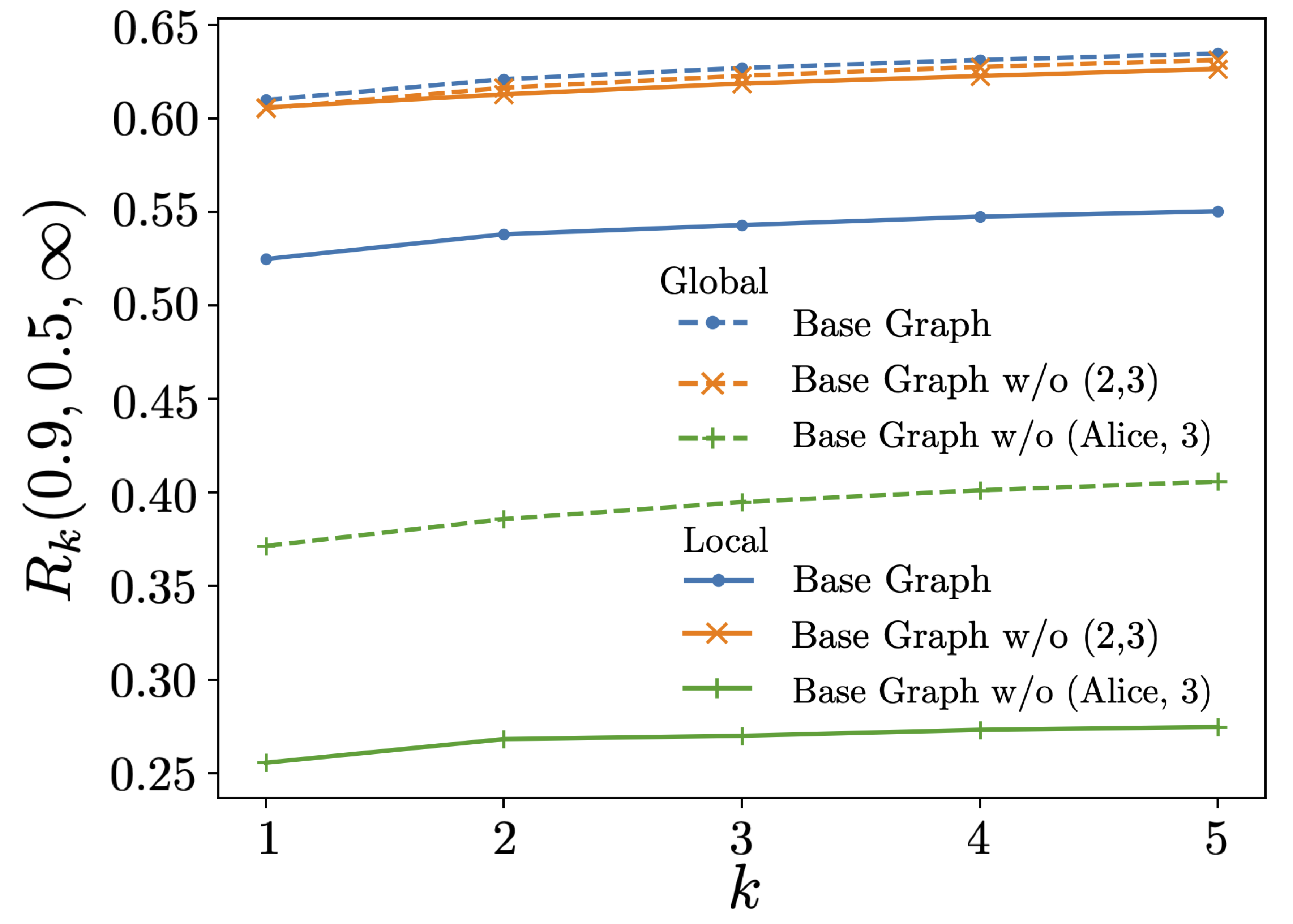}
    \caption{The average rates found using global knowledge (dashed lines) and local link state knowledge (solid lines) are depicted for three variations of the six-node network shown in Fig.~\ref{fig:6node}. Base Graph refers to this network with all channels present. Removal of certain channels of this base graph is shown to always hurt the average rate of the global algorithm, but can improve the average rate of the local algorithm.}
    \label{fig:6nodeRate}
\end{figure}

In this section, we evaluate the performance of the protocols on the six node network depicted in Fig.~\ref{fig:6node}. This network is representative of near-term quantum networks, which will have a small number of nodes that span across a building, a campus, or an urban center. The size of the network was also chosen so that the average rate of the dynamic local link state knowledge protocol could be compared against two global link state knowledge routing protocols introduced in \cite{pant_routing_2019} and \cite{vardoyan2023bipartite}.

For the dynamic local link state knowledge protocol, shown in Fig.~\ref{fig:6nodeRate} by the solid lines, the distance metric utilized to inform decisions was based on hop distance. Two different calculations were done to obtain the global link state knowledge rates. Firstly, the mixed-integer quadratically constrained program (MIQCP) code from \cite{vardoyan2023bipartite} was used to calculate the maximum rate  for every potential snapshot, or unique instance, of the network. The average rate of the network, provided global link state knowledge and optimal decision making, was then calculated by summing over the snapshot rates weighed by their likelihood of occurring. This was then compared against a suboptimal scheme presented in \cite{pant_routing_2019}. The main difference was the latter protocol used global link state knowledge and repeatably applied Dijkstra's algorithm to find the rate of each snapshot. The rates found by this greedy algorithm, which may not be optimal, were similarly weighed by their likelihood, providing the average rate with global link state knowledge and sub-optimal decision-making. The benefit of such a protocol is that the calculation can be performed far more quickly. These results were in agreement up to seven significant digits, and are represented by the dashed lines in Fig.~\ref{fig:6nodeRate}.

Three versions of the six node graph shown in Fig.~\ref{fig:6node} were evaluated. The first included all the channels present in the figure. The second included all, but the channel between repeater 2 and repeater 3. The third included all the channels, but the channel between Alice and repeater 3. The average rates given by the global algorithms did best when more channels were present, since there were simply more potential paths that could be chosen to route over. However, the local protocol did best when the channel between repeaters 2 and 3 was removed. Similar to Braess' paradox in classical routing, where adding more roads to a road network can slow down the overall traffic, our local algorithm can perform worse on networks when there are more channels if these channels create a contention for paths. Consider a snapshot where all links succeed except for the one between Alice and repeater 3.  Since the repeater nodes are making decisions on their own, the local protocol can result in the path given by Alice, 2, 3, Bob, which  intersects the two otherwise disjoint paths. The dynamic protocol performs best on topologies where there are multiple subsections of the graph connected in parallel at the consumer nodes. Otherwise, fixed path routing or some form of network-divisions similar to those introduced in \cite{patil_entanglement_2022} should be used to work around this.

\subsection{\label{sec:2D Square Lattice}2D Square Lattice}

\begin{figure}
    \centering
    \includegraphics[width=.45\textwidth]{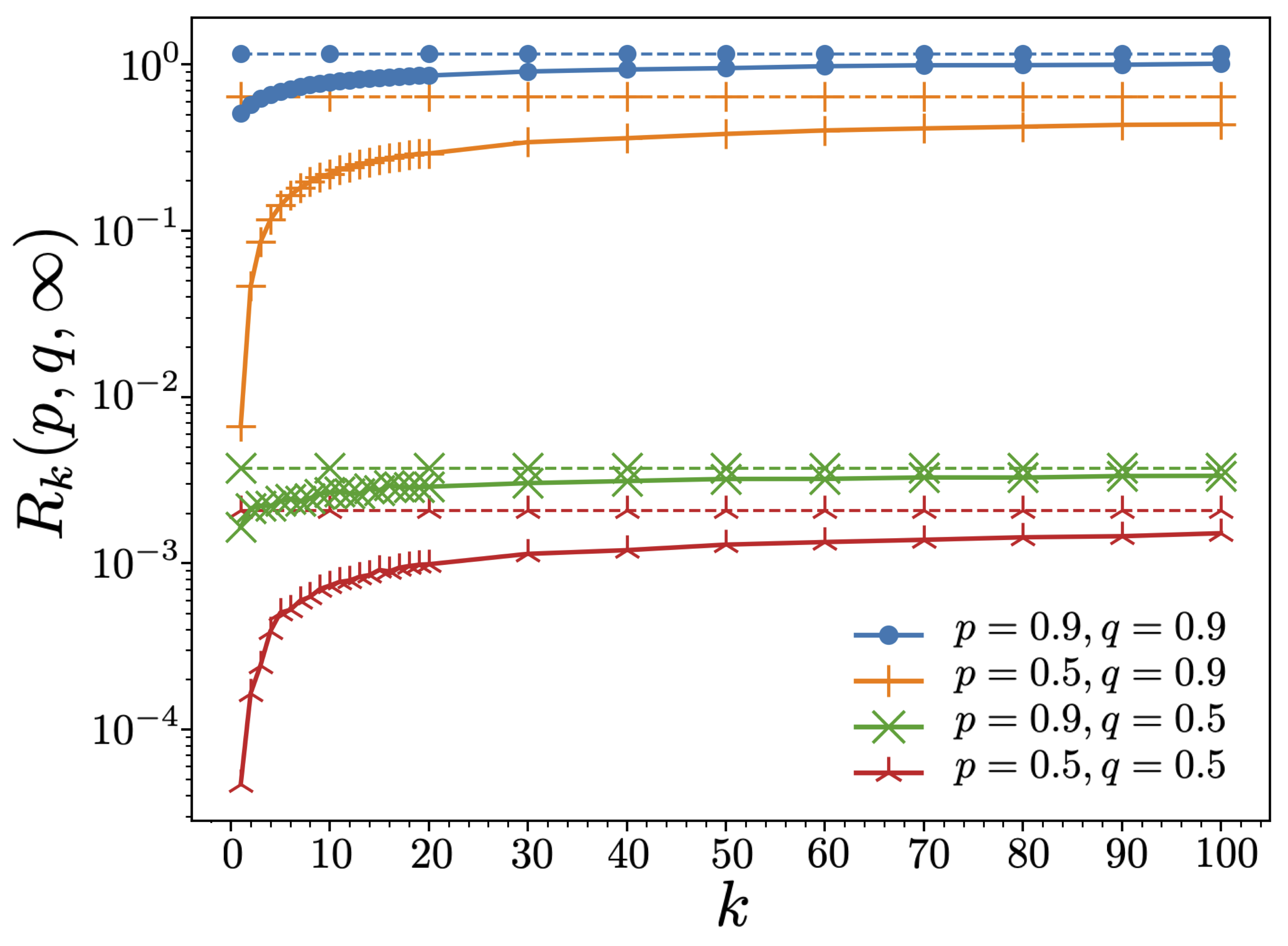}
    \caption{The average rate of the dynamic protocol, indicated by solid lines, approaches the values predicted for $R_{\infty}(p,q,\infty)$. The corresponding bounds are shown above, indicated by dashed lines, for the case that Alice and Bob were both along the diagonal with a hop distance of 10 between them. Low values of $q$ are more detrimental to the rate, as the dynamic routing protocol and time multiplexing can compensate for low $p$ values.}
    \label{fig:No Dec, diff pq}
\end{figure}
\begin{figure}
    \centering
\includegraphics[width=.45\textwidth]{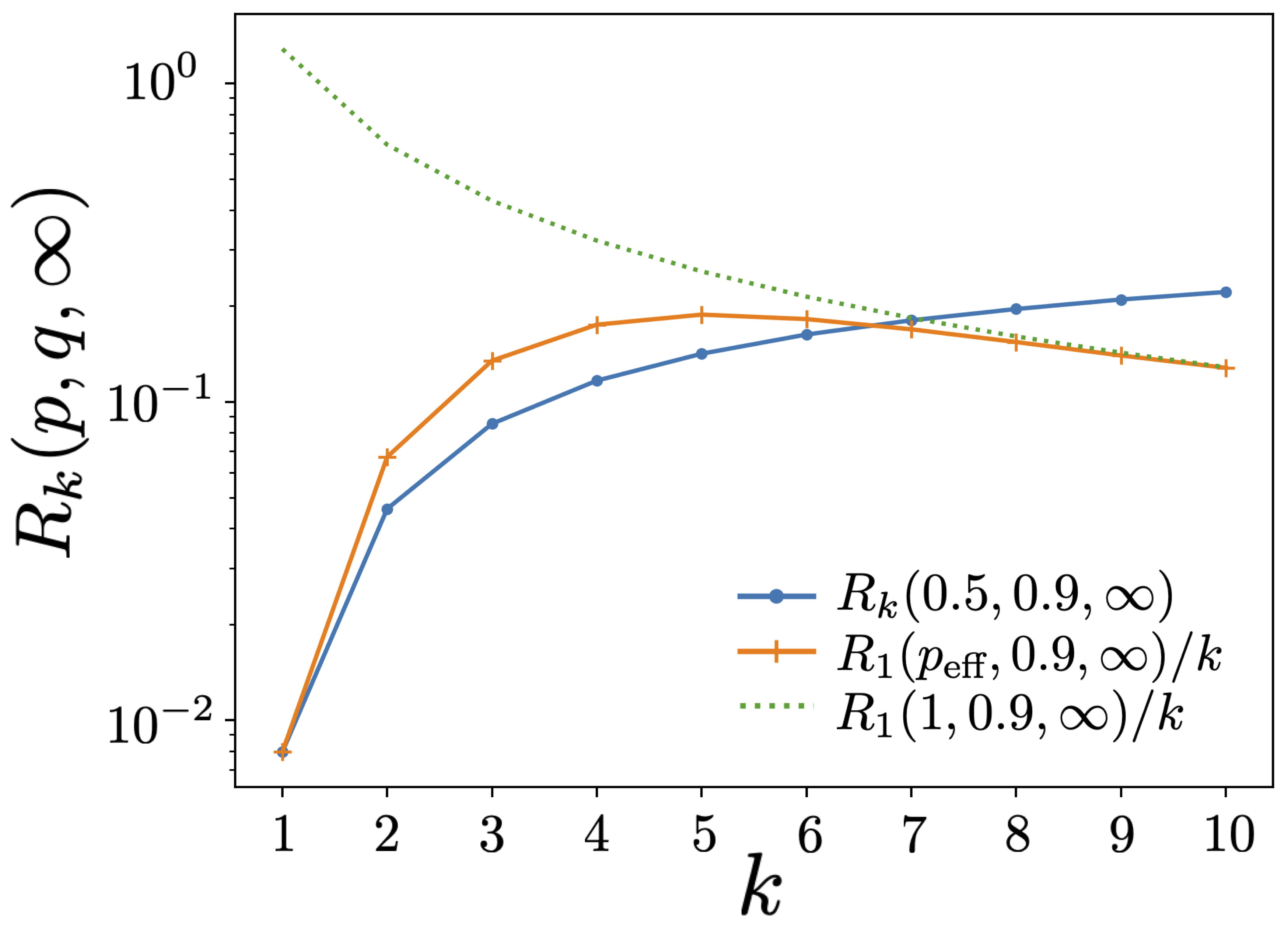}
    \caption{Comparison of the average rate achievable with time multiplexing when using only one success per edge vs using multiple successes across timesteps.  $R_1(p_{\text{eff}},0.9,\infty)/k$, when $p=0.5$, shows the average rate when limited to only using one success. $R_1(1,0.9,\infty)/k$ provides an upper bound to this and is included to to highlight the inverse scaling with $k$. The average rate when you are allowed to use multiple successes is given by $R_k(0.5,0.9,\infty)$. Data was collected from a 2D square grid where Alice and Bob were 10 hops away apart and on the diagonal.}
    \label{fig:1succ}
\end{figure}
\begin{figure}
    \centering
    \includegraphics[width=.45\textwidth]{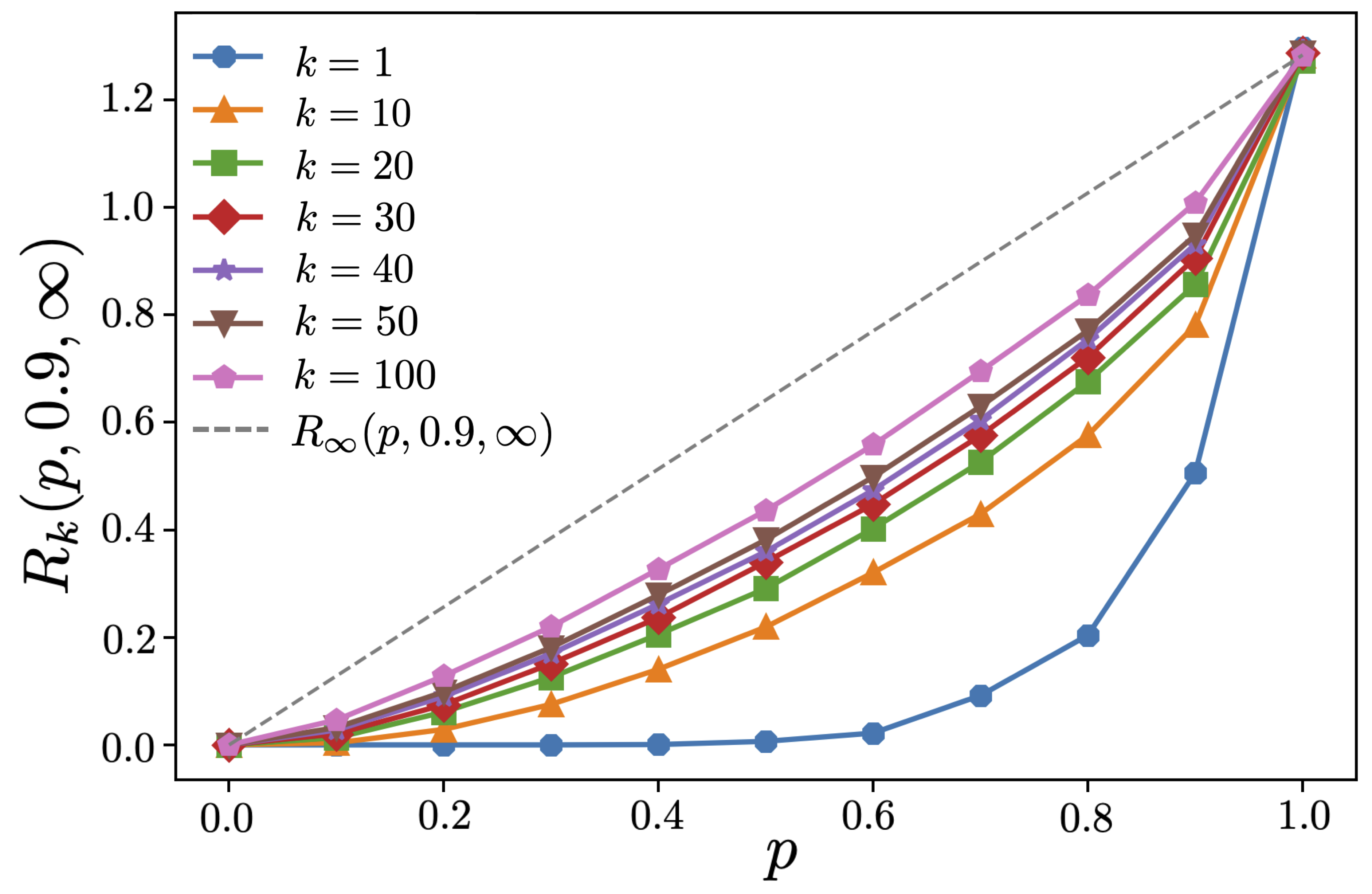}
    \caption{As $k$ increases, the average rate of the dynamic protocol approaches the predicted bound, indicated by the dashed line. As $k\xrightarrow{}\infty$, a linear relationship emerges between the average rate and $p$. The data depicted here is for the case where Alice and Bob were both along the diagonal with a hop distance of 10 between them.}
    \label{fig:No Dec, R vs p}
\end{figure}

\begin{figure}
    \centering    \includegraphics[width=.45\textwidth]{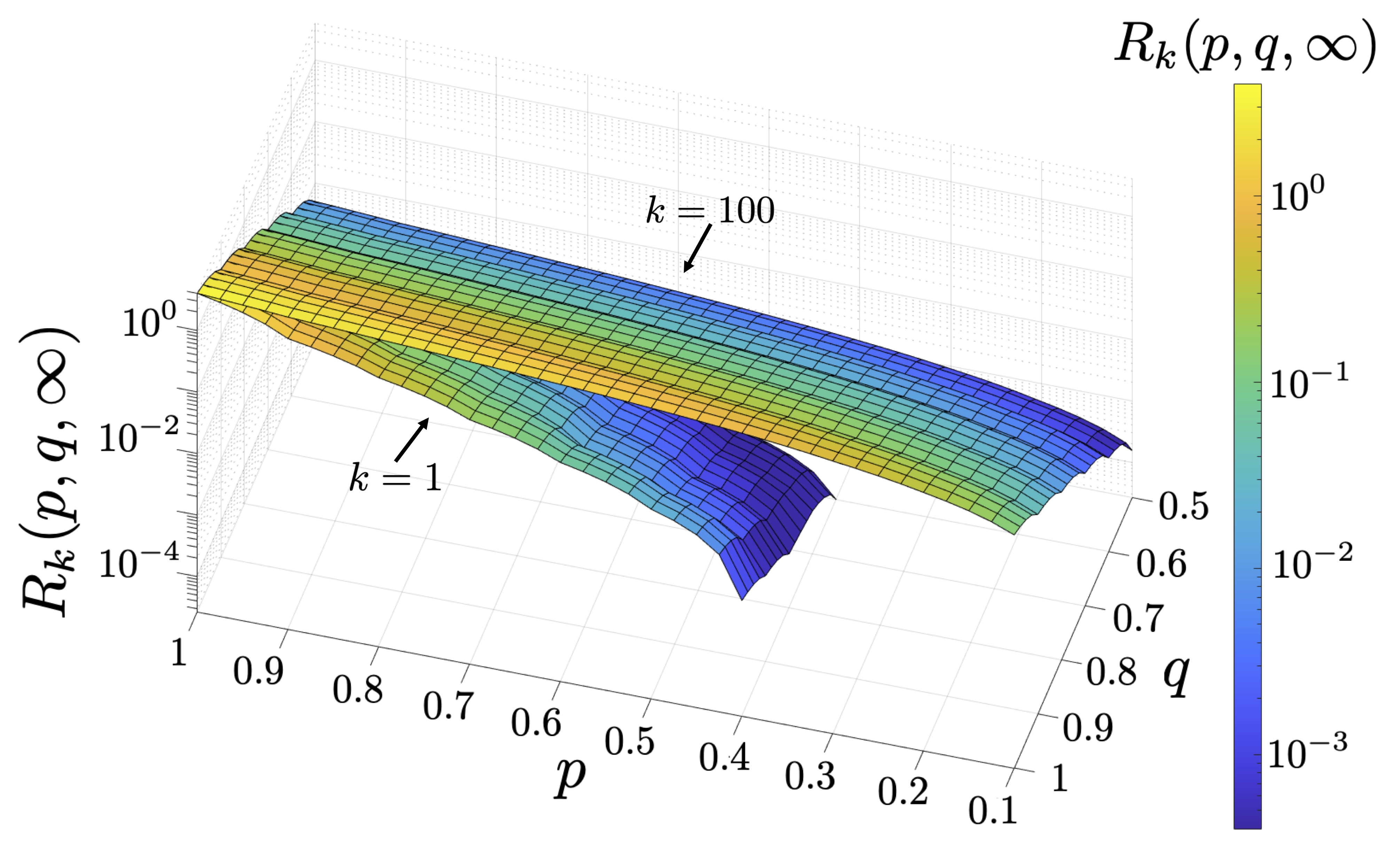}
    \caption{Given that Alice and Bob are both along the diagonal with a hop distance of 10 between them, the average rates for $k=1$ and $k=100$ are compared over different values of $p$ and $q$. Assuming there is no decoherence, increasing the time multiplexing block length can only help the average rate.}.
    \label{fig:No Dec, k=1 vs k=100}
\end{figure}
\subsubsection{Initial Performance Evaluation}

Two consumers placed on a square grid can have at most four edge disjoint paths connecting them. When time multiplexing is introduced so that any edge can have up to $k$ links, the maximum number of connections between consumers is $4k$. If $k$ is sufficiently large, the average number of links along a given edge will be $pk$. Letting $q=1$, we expect $4pk$ connections between consumers. Dividing this by $k$ yields an average rate of $4p$ Bell pairs per time slot. For general $q$, in the limit as $k$ gets large, the highest that the average rate can get is given by
\begin{align}
\label{Bound}
R_{\infty}(p,q,\infty)=pR_{1}(1,q,\infty)\nonumber\\
=p\sum_{i=1}^{\theta}q^{m_i-1}
\end{align}
 where $m_i$ is the hop distance of the $i$th shortest path on the underlying graph, which is found via the greedy algorithm.

As depicted in Fig.~\ref{fig:No Dec, diff pq}, the dynamic protocol is able to approach this bound on the square grid. We note that low values of $q$ are more detrimental to the rate than low values of $p$. This is because time multiplexing can compensate for low $p$ values. The probability that an edge has at least one successful link after $k$ timesteps is 
\begin{align}
\label{p_eff}
p_{\text{eff}}=1-(1-p)^k,
\end{align} 
which monotonically increases with $k$. If we were to use only one success along any edge, the rate would then be equivalent to $R_1(p_{\text{eff}},q,\infty)/k$, which scales as $k^{-1}$. However, the dynamic protocol utilizes as much entanglement as possible, ultimately overcoming this scaling as shown in Fig.~\ref{fig:1succ}. When $k$ is small, there are cases where the opposite is true. In Fig.~\ref{fig:1succ}, when $k\in[2,6]$, using multiple successes results in a lower average rate because there is an increased likelihood of paths to self loop. The value of $p$ determines how quickly the average rate approaches $R_{\infty}(p,q,\infty)$, as shown in Fig.~\ref{fig:No Dec, R vs p}.

When $p=1$, time multiplexing doesn't show an improvement to rate since there is no improvement to $p_{\text{eff}}$, as shown in Fig.~\ref{fig:No Dec, k=1 vs k=100}. It is important to note that while the average rate only benefits from larger $k$, the corresponding latency suffers. Therefore realistically, one should choose some finite $k$ based on the desired latency.

\subsubsection{Dynamic versus fixed path routing}
We now compare the dynamic protocol against the static protocol and determine the preferred protocol based on the network parameter regime. Generally, for the 2D square grid graph the distance based routing protocol does best when a Euclidean distance metric is used to make swapping decisions, as opposed to a hop distance metric. However, even with this metric, the dynamic protocol is only able to outperform the static protocol in certain situations.

When $k=1$ and the consumers are oriented along the diagonal, the dynamic protocol outperforms the static protocol for $p<1$. For this specific orientation, the dynamic protocol naturally maps over the same paths as those chosen by the static paths algorithm, unless those links do not exist. However, unlike the static protocol, the dynamic protocol is able to adapt when there are failed links. When $p=1$, the static and dynamic protocols perform equally well. This behavior is consistent regardless of the distance between consumers.

As $k$ increases however, a region emerges in $p, q$ space, shown in Fig.~\ref{fig:No Dec, Static vs Dyn}, where the static protocol outperforms the dynamic protocol. This behavior stems from the additional ordering that the static path protocol introduces, so that it is able to more efficiently use successful entangled links than the dynamic protocol. The static path protocol does better when $p$ is large, but the exact size and shape of this region depends on $k$ and the distance between consumers. As the distance between consumers grows, the static protocol performs better for lower $p$ values. 

When $k=1$ and the consumers are not oriented along the diagonal, there already exists a region in $p,q$ space where $p$ is large where the static protocol outperforms the dynamic protocol. This is because when consumers aren't along the diagonal, the dynamic protocol doesn't always choose the most efficient paths. Thus there exists a value $p$ that is large enough that the ability to reroute paths is less useful than the directness of the paths provided by the static path protocol. When $q=1$ the length of a path no longer matters, as all BSMs are deterministic. However, there still exists a range of values of $p$ where the static protocol does better, showing that the dynamic protocol not only chooses longer paths, but also makes decisions that do not result in continuous paths between consumers. As $k$ increases, the quantity $p_{\text{eff}}$ defined in  \eqref{p_eff} increases, explaining why the borders shift to lower $p$ values.

 When $k>1$ and the Manhattan distance is fixed, the border separating the regions where the dynamic protocol does better from where the static protocol does better occurs at lower $p$ values for the column orientation and higher $p$ values for diagonal orientation. The border for other angles falls between these two cases.

When $k>1$ and the angle is fixed, the border shifts to lower $p$ values as the Manhattan distance is increased. The larger distance between consumers requires more successful BSMs. The ability to reroute is not as important, so static paths do better since they are chosen to minimize the number of these BSMs. As $q$ goes to one however, the difference in path lengths does not matter as much, explaining why the borders appear to tilt diagonally upwards.

The network topology and the relative locations of the consumers determine the effectiveness of the dynamic protocol. This section shows that the effectiveness of the dynamic distance based routing algorithm is impacted by the network topology as well as the consumer locations.  The dynamic protocol is the better option when networks have low connectivity, as it can adaptively choose different repeaters to route over. However, static or fixed path routing is usually preferable when connectivity is high, which we also saw in the case of the small network example. Again we conclude that additional topology related constraints should be incorporated in our dynamic protocol to combat this effect.

\begin{figure}
    \centering
\includegraphics[width=.45\textwidth]{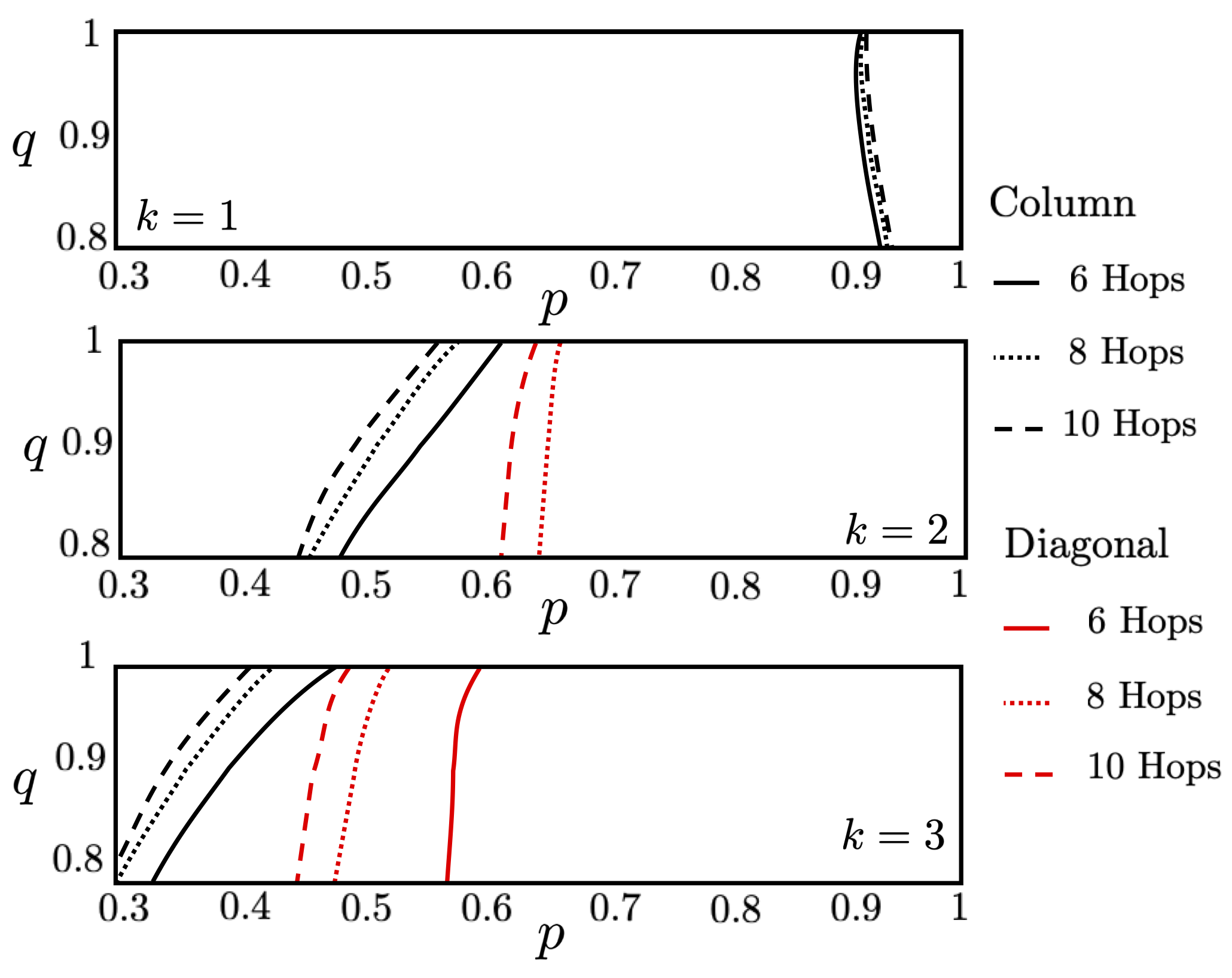}
    \caption{The regions where the dynamic/static protocol performs better for consumers along the column and diagonal with different hop distances apart are demarcated. To the left of the lines the dynamic protocol does better and to the right of the lines the static protocol does better. As $k$ increases, the range of $p$ and $q$ values where the dynamic protocol outperforms the static diminishes. If a line is not shown on the plot, the dynamic protocol does better for all $p,\ q$ values.}
    \label{fig:No Dec, Static vs Dyn}
\end{figure}

\section{\label{sec:Results with imperfect quantum memories} Memories with limited storage lifetime}
The previously discussed results imply that as long as consumers are willing to wait for some increased latency cost, time multiplexing always helps improve the rate of entanglement distribution. However, as time multiplexing blocks increase in length, qubits stored in the quantum memories experience more decoherence. We now relax the assumption that qubits are held in the quantum memories indefinitely and instead will evaluate for a finite valued $\mu$. This model will be applied to a 2D grid network to show that for given network conditions, there arises a finite $k$ for which the rate of entanglement is maximized.

\begin{figure}
    \centering
    \includegraphics[width=.45\textwidth]{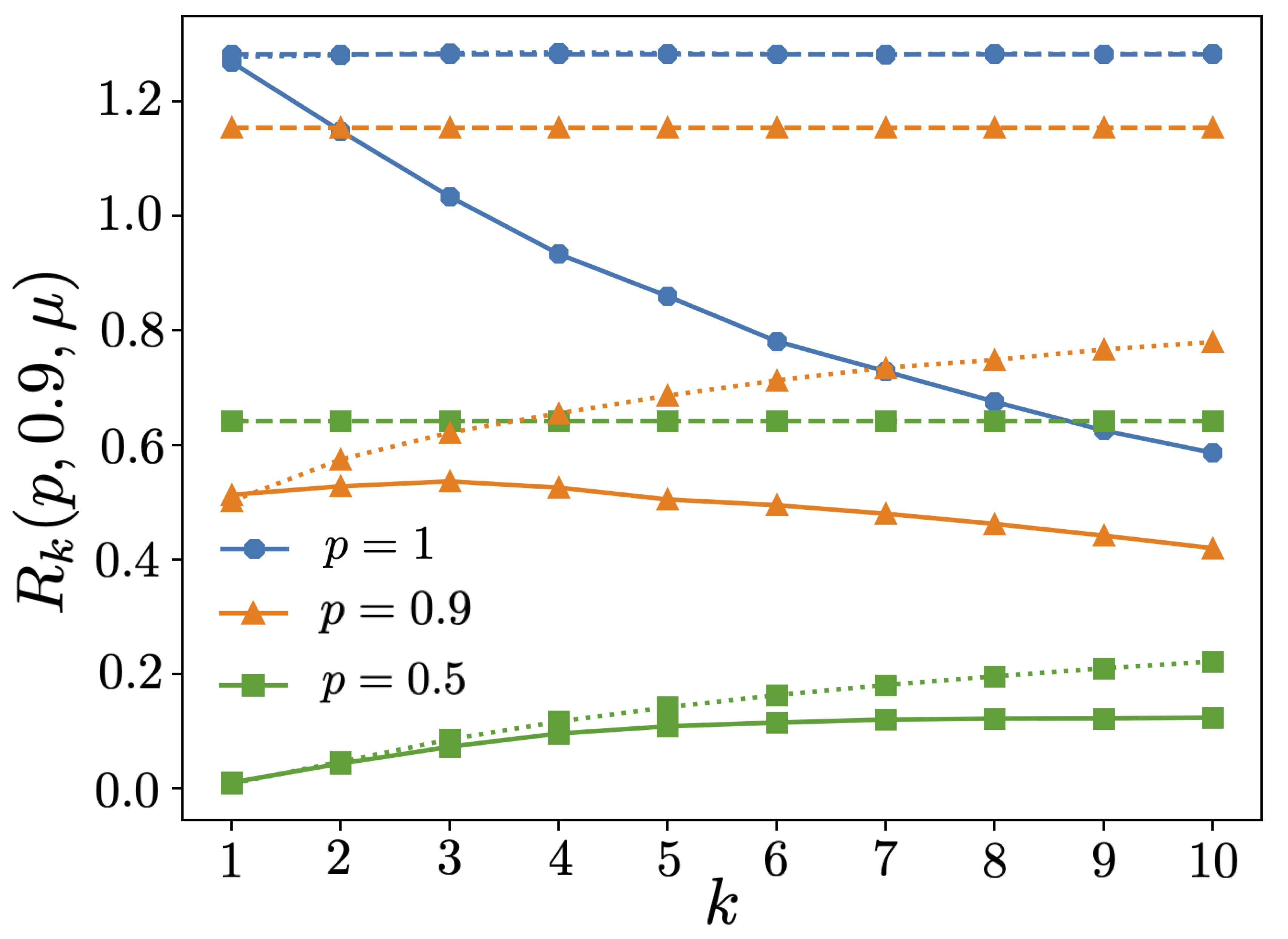}
    \caption{The average rates for a variety of $(p,q)$ values are shown here in the presence and absence of decoherence. The solid line is for $\mu=100$, the dotted line is for $\mu=\infty$ and the dashed line shows the upper limit $R_\infty(p,q,\infty)$. In the case of $p=1$, the dotted and dashed lines overlap.}
    \label{fig:DecvsNoDec}
\end{figure}

\begin{figure}
    \centering
    \includegraphics[width=.45\textwidth]{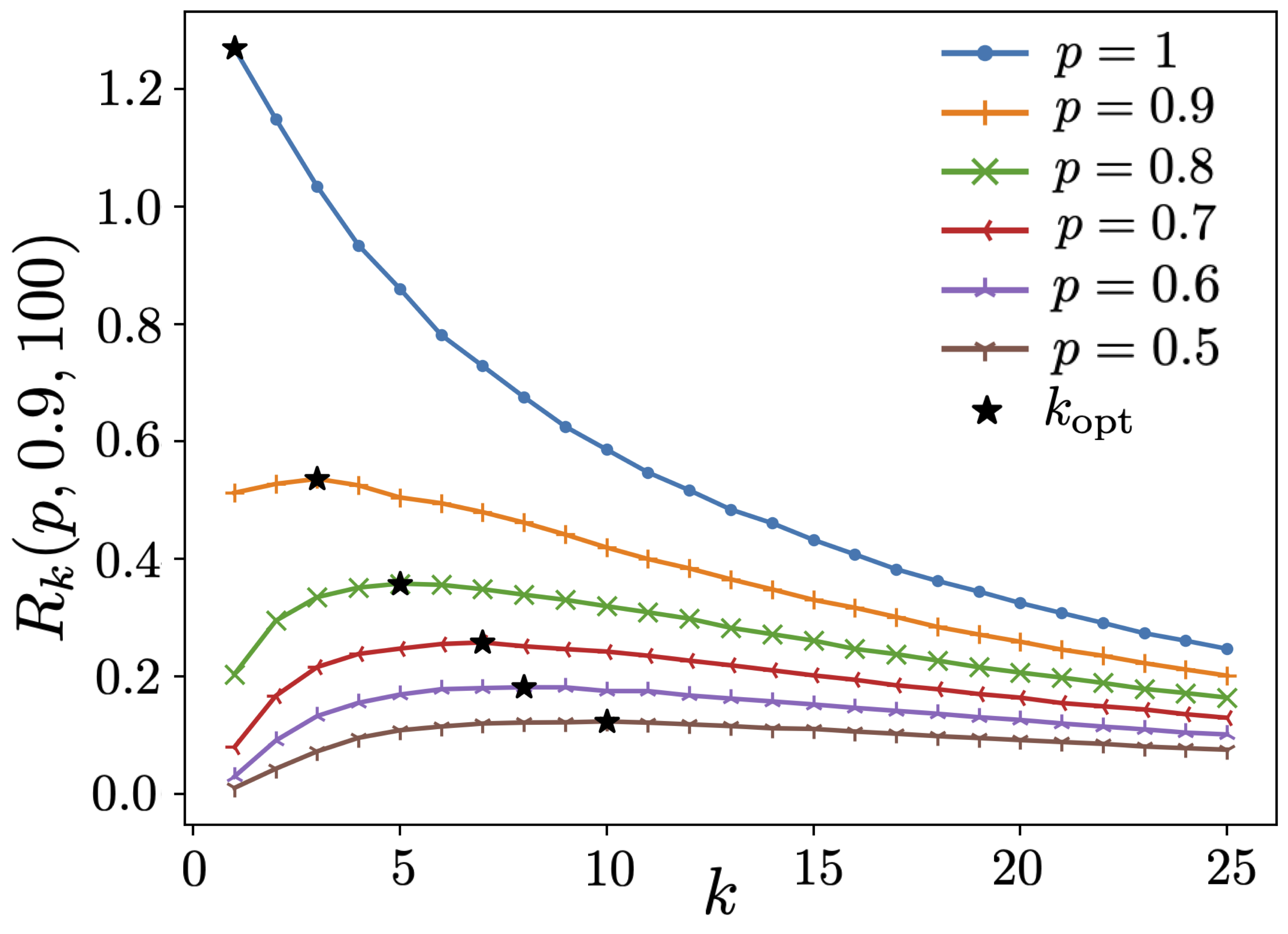}
    \caption{In the presence of decoherence, each value of $p$ has an optimal value $k$, denoted as $k_{\text{opt}}$, such that the average rate is largest for that time multiplexing block length. As $p$ decreases, $k_{\text{opt}}$ increases.}
    \label{fig:With Dec, R vs k diff p}
\end{figure}

\begin{figure}
    \centering
    \includegraphics[width=.45\textwidth]{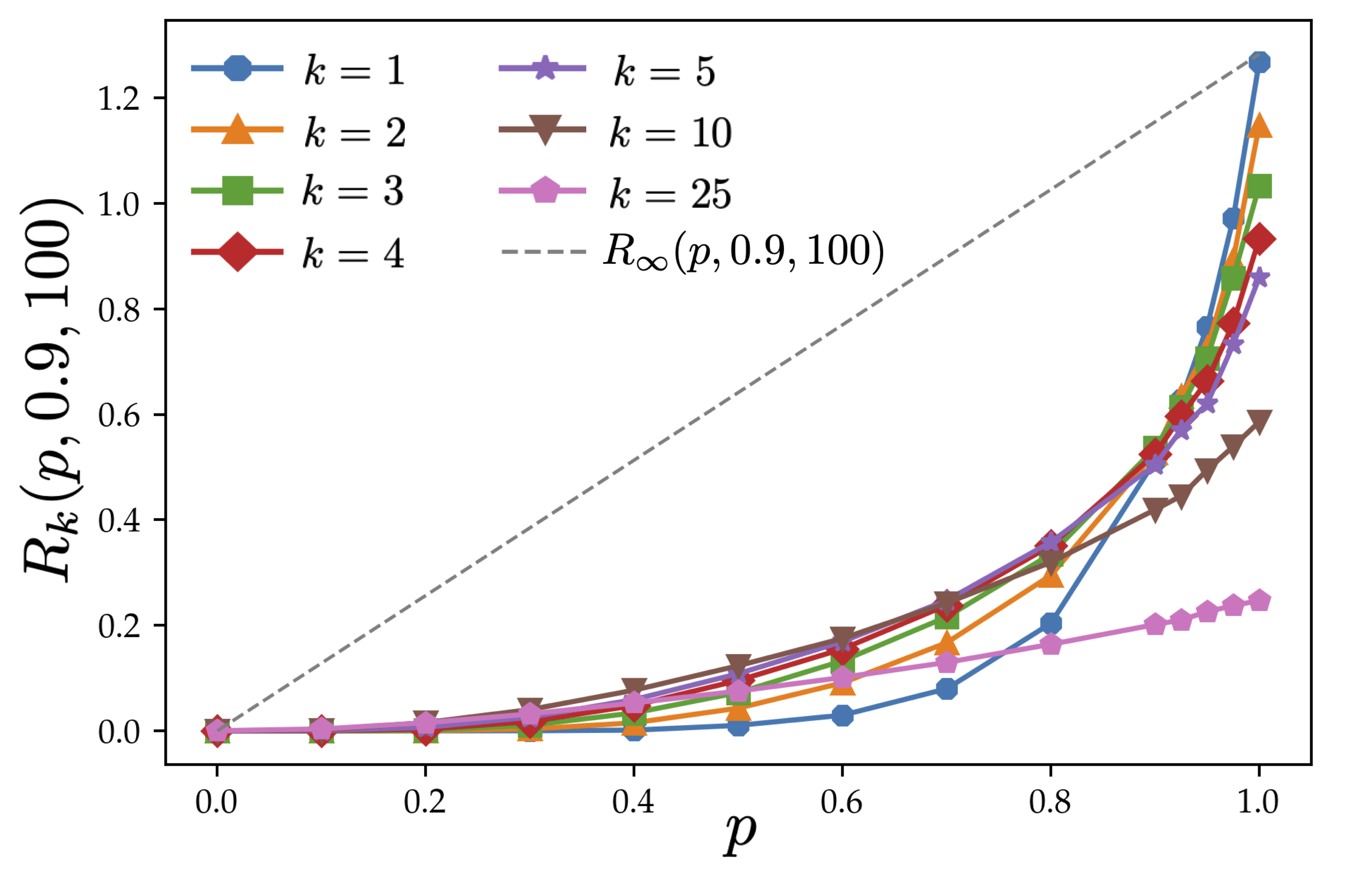}
    \caption{In the presence of decoherence, time multiplexing can help bridge the gap between $R_1(p,q,100)$ and $R_{\infty}(p,q,\infty)$. However, this improvement is bounded by an envelope determined by the value of $\mu$.}
    \label{fig:With Dec, R vs p}
\end{figure}

\begin{figure}
    \centering
    \includegraphics[width=.45\textwidth]{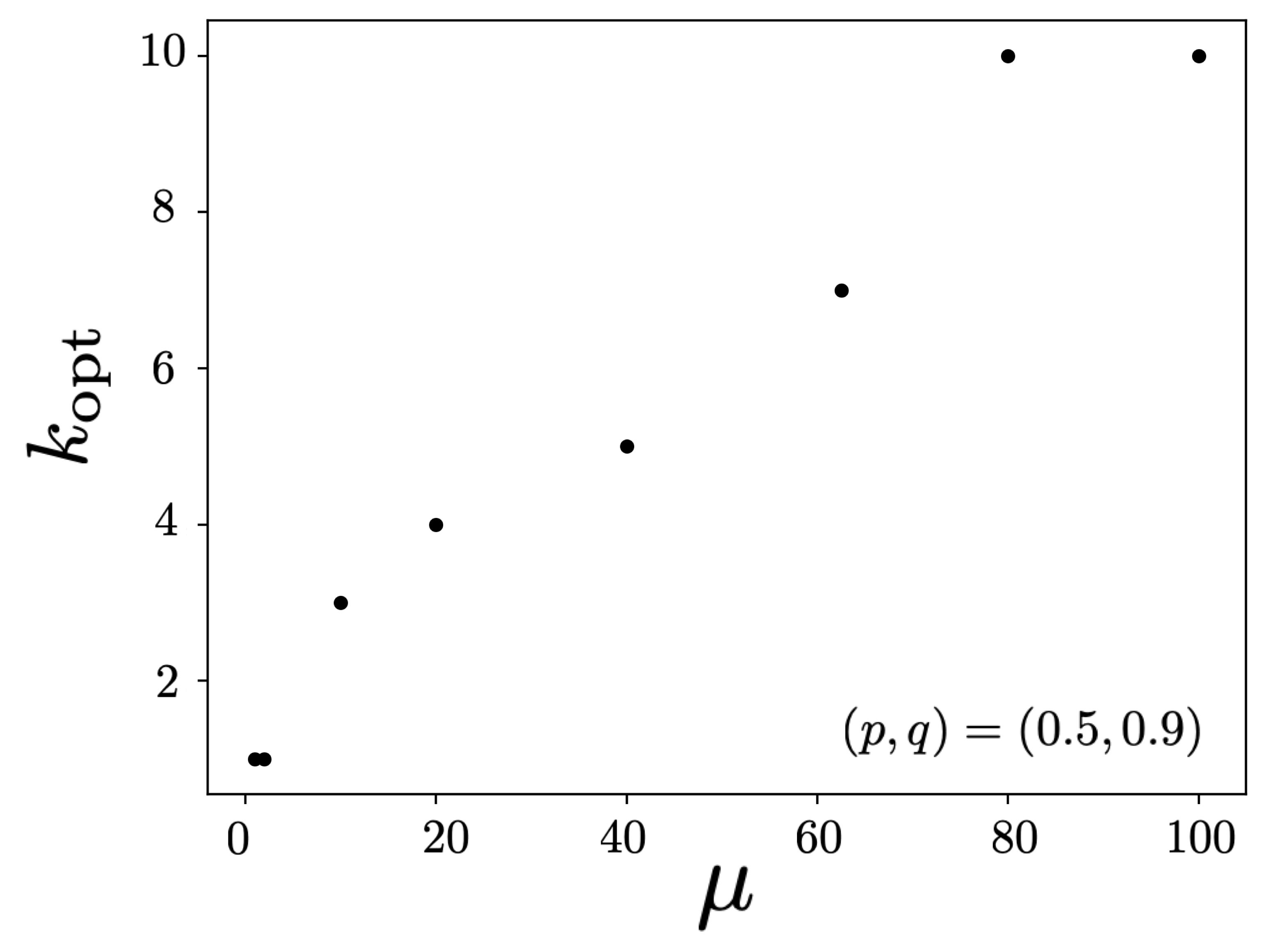}
    \caption{There is a monotonically increasing relationship between 
    $k_{\text{opt}}$ and $\mu$, which in the limit of $\mu\xrightarrow{}\infty$ agrees with the results without decoherence. Note, $k=10$ was the maximum value tried for $k$.} 
    \label{fig:With Dec, k vs mu}
\end{figure}

\begin{figure}
    \centering
\includegraphics[width=.45\textwidth]{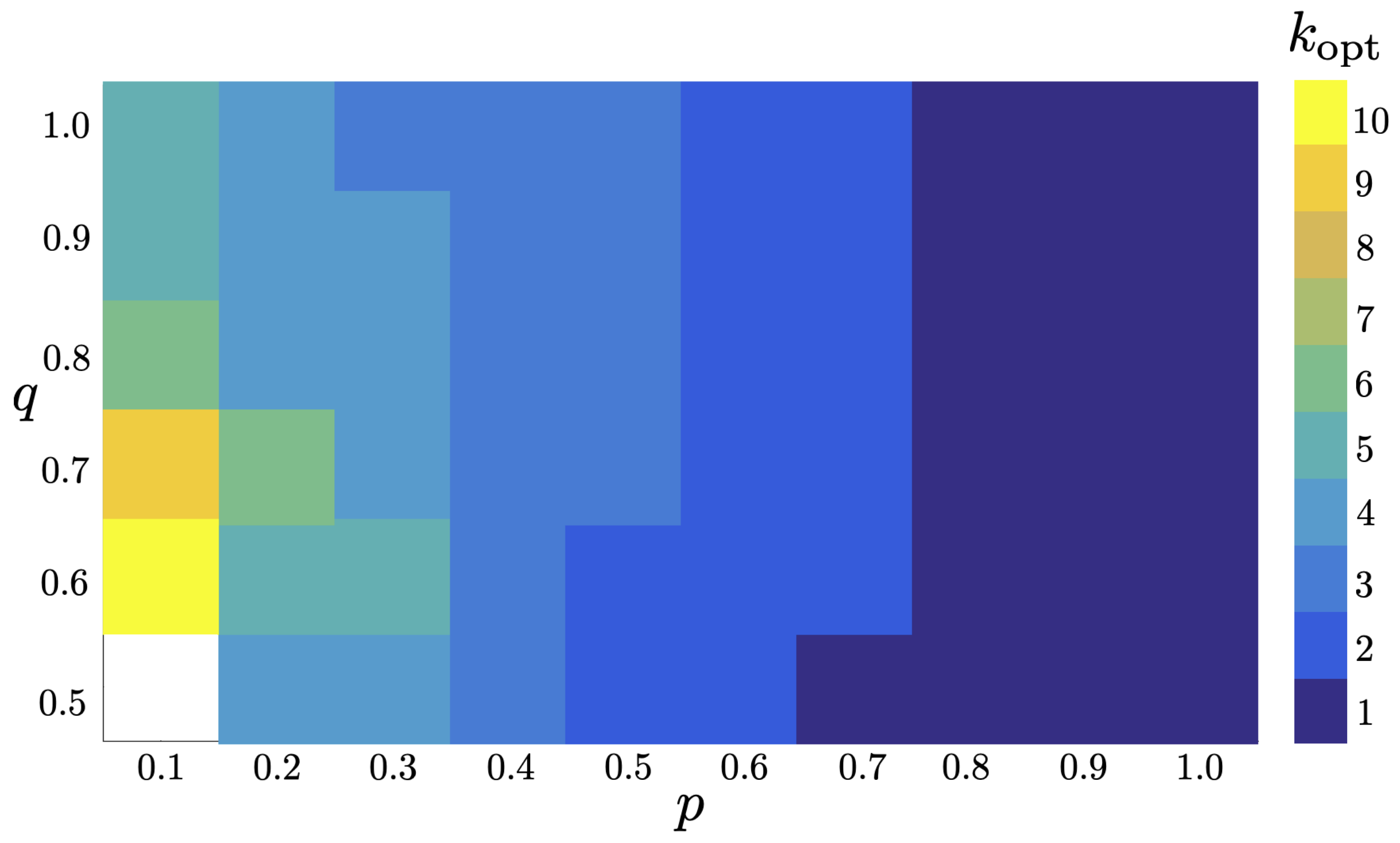}
    \caption{The value of $k_{\text{opt}}\in[1,10]$ is indicated here for a range of values of $p$ and $q$. No optimal k was found in the white region, as the rate remained zero over all trials. }
    \label{fig:With Dec, Optimal k}
\end{figure}

\begin{figure}
    \centering
\includegraphics[width=.45\textwidth]{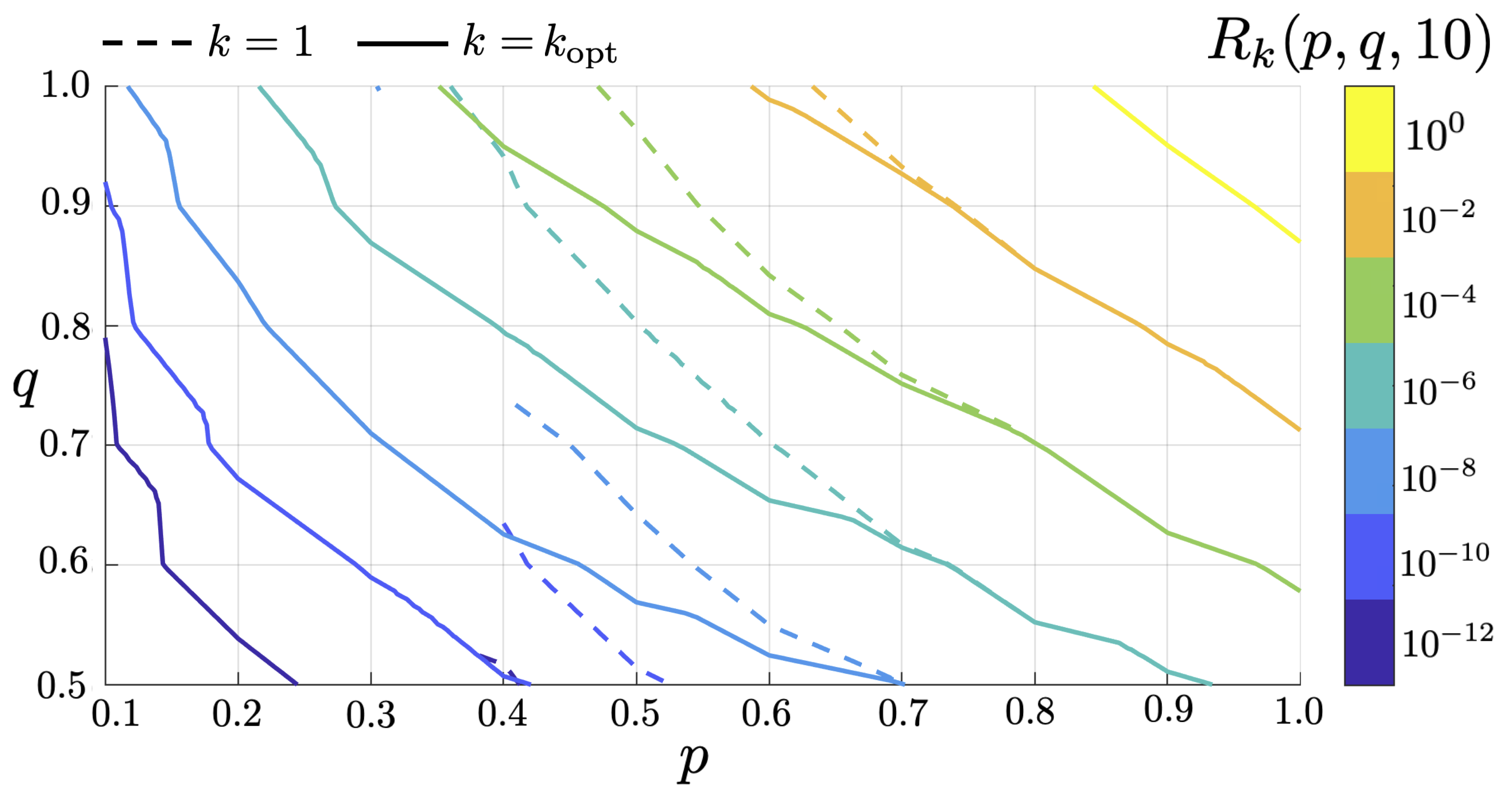}
    \caption{The solid lines here indicate the contour lines of $R_{k_{\text{opt}}}(p,q,10)$ whereas the dashed lines demarcate the contour lines of $R_{1}(p,q,10)$. Note the dashes lines do not span the entire space, because due to the finite number of trials run, they returned rates of zero below $p=0.4$.}
    \label{fig:With Dec, Optimal Rate}
\end{figure}

For such a network, there exists a trade-off between the benefits of increased time multiplexing block length and the risk of losing already established entangled links, as shown in Fig.~\ref{fig:DecvsNoDec}. This relationship results in an optimal time multiplexing value, $k_{\text{opt}}$, dependent on the network conditions. Fig.~\ref{fig:With Dec, R vs k diff p} shows that as $p$ decreases, $k_{\text{opt}}$ increases. When $p$ goes to one, so that external links are deterministic, time multiplexing is no longer useful, and $k_{\text{opt}}=1$. Fig.~\ref{fig:With Dec, R vs p} shows that even with decoherence, there is an improvement in rate with time multiplexing, as seen by the envelope. The amount of improvement is limited by the magnitude of $\mu$. As $\mu$ goes to infinity, the achievable rates with time multiplexing will increase, eventually resembling Fig.~\ref{fig:No Dec, R vs p}. Generally, increasing $\mu$ increases  $k_{\text{opt}}$, as seen in Fig.~\ref{fig:With Dec, k vs mu}. In the case where quantum memories have infinite lifetime, and thus no decoherence, the rate continues to increase with $k$. Fig.~\ref{fig:With Dec, Optimal Rate} shows the improved rate over $(p,q)$ space, using the values of $k_{\text{opt}}$ shown in Fig.~\ref{fig:With Dec, Optimal k}. There is noise in the data stemming from the finite number of trials the Monte Carlo simulation was run over and the limited data points collected, so it is not immediately clear if the value of $q$ impacts the optimal $k$, despite it clearly affecting the average rate. However, it is once again clear that time multiplexing helps, especially in the case of lower $p$.

\section{\label{sec:Conclusion}Conclusion}
This study discusses a local link state knowledge multipath routing protocol that utilizes time multiplexed repeaters. In the case of no decoherence, the average entanglement rate increases monotonically with the time multiplexing block length $k$. This is because the probability of have a link between repeaters increases with $k$. Theoretically, to maximize the rate of entanglement distribution, it is best to let $k\xrightarrow{}\infty$. This is not practical, however, as latency and the required number of quantum memories also become infinite. Therefore, the value of $k$ must be chosen while keeping both the initial latency and memory buffer length in mind. Future work should consider the effects of having a fixed number of quantum memories in the repeaters on optimal scheduling.

The distance metric chosen and the topology of the network itself, impacts how effectively the dynamic protocol performs with respect to a static fixed path routing algorithm. For the case of a 2D grid, the dynamic protocol performs best when consumers reside on the diagonal. In general, the performance of the dynamic protocol when compared to the static protocol depends on the values of $p,\ q,\ k$ and the locations of the consumers. The dynamic protocol was also applied to a smaller six node network, so that it could be compared to the global optimum rate. The local link state knowledge protocol can perform comparably well when consumers are connected via disjoint subsets of repeaters. However, when there is a high contention for the same paths, our local protocol cannot approach the average rates seen with global link state knowledge. Future topology aware routing protocols
should be explored to help improve upon the performance of our two local knowledge based protocols.

A step function decoherence model was also used in the simulation to see how the mean lifetime of qubits held in the quantum memories impacted the effects of time multiplexing. In this case, an optimal $k$ value emerges, which balances the benefits from time multiplexing with the increased risk of losing a Bell pair. It was seen that as $p$ decreases or $\mu$ increases, the value of $k_{\text{opt}}$ increases. To fully understand the effect of $q$ on $k_{\text{opt}}$, further study is needed; however, we conjecture that there may be a connection stemming from the fact increasing $k$ allows for more direct paths to be routed over. Overall, the results with decoherence agree with the results without decoherence when $\mu\xrightarrow{}\infty$.

The decoherence model used in this paper can be improved upon by introducing non-unit fidelity links. These links could then undergo depolarization noise channels to model both the effects of being held in the quantum memory over time, as well as the noise introduced by the imperfect gates used to perform BSMs.  With this, time multiplexing could then be combined with distillation techniques, which would result in purer states, but at the cost of reducing to the number of Bell pairs delivered to the consumers over some fixed time. A protocol dictating how and when to distill links would then need to be considered. 

\section*{Acknowledgment}
EV thanks Prithwish Basu for useful discussions. This material is based upon High Performance Computing (HPC) resources supported by the University of Arizona TRIF, UITS, and Research, Innovation, and Impact (RII) and maintained by the UArizona Research Technologies department.

\bibliographystyle{IEEEtran}
\bibliography{refs}

\appendices
\section{\label{psuedocode}Distance Based Routing Pseudocode}
In this appendix we provide the pseudocode for the External and Internal Phases of the Distance Based Routing Protocol.

\begin{algorithm}[H]
  \caption{External Phase}
  \label{External Phase}
   \begin{algorithmic}[1]
       \For{each edge $e$ in physical network}
       \State{$n_1=\ $source($e$)}
       \State{$n_2=\ $target($e$)}
        \For{each timestep $t$ from 1 to $k$}
       \If{UnifRand$(0,1)<p$}\Comment{Link success}
        \If{ExpRand(1/$\mu)\geq k-t$}\Comment{Decoherence}
        \State{AddLink($n_1,n_2$)}
        \EndIf
        \EndIf
        \EndFor
        \EndFor
   \end{algorithmic}
\end{algorithm}
\begin{algorithm}[H]
  \caption{Internal Phase - Distance Based Routing }
  \label{Internal Phase}
   \begin{algorithmic}[1]
       \For{each node $n$ in physical network}
       \State{ExtLinks = GetLinks($n$)}\Comment{Lists external links at $n$}
       \State{$s=\ $len(ExtLinks)}\Comment{Number of external links at $n$}
        \While{$s>1$}
        \State{$v$ $\gets$ closest neighbor to Alice}
        \State{$w$ $\gets$ closest neighbor to Bob}
       \If{$v\neq w$}
       \State{$m_v=$ Qubit$(n,v)$}\Comment{memory at $n$ linked to $v$}
       \State{$m_w=$ Qubit$(n,w)$}
       \ElsIf{$\exists$ links between $n$ and neighbor $\neq v$}
        \State{$v'$ $\gets$ second closest neighbor to Alice}
        \State{$w'$ $\gets$ second closest neighbor to Bob}
        \If{$d_{Av'} + d_{Bw} < d_{Av} + d_{Bw'}$}
        \State $m_v=$ Qubit$(n,v')$
        \State $m_w=$ Qubit$(n,w)$
        \ElsIf{$d_{Av'} + d_{Bw} > d_{Av} + d_{Bw'}$}
        \State $m_v=$ Qubit$(n,v)$
        \State $m_w=$ Qubit$(n,w')$
        \ElsIf{$d_{Bv'} + d_{Aw} > d_{Bv} + d_{Aw'}$}
        \State $m_v=$ Qubit$(n,v')$
        \State $m_w=$ Qubit$(n,w)$
        \Else
        \State $m_v=$ Qubit$(n,v)$
        \State $m_w=$ Qubit$(n,w')$
        \EndIf
        \Else\Comment{Form a self connection}
        \State $m_v=$ Qubit$(n,v)$
        \State $m_w=$ Qubit$(n,w)$
        \EndIf
        \If{UnifRand$(0,1)<q$}\Comment{BSM success}
        \State{BSM($m_v,m_w$)}

        \EndIf
        \State{ExtLinks.remove(Link($m_v$))}
        \State{ExtLinks.remove(Link($m_w$))}
        \State{$s=s-2$}
        
        \EndWhile
        \EndFor
   \end{algorithmic}
\end{algorithm}

\end{document}